\documentclass[journal= langd5,manuscript=article]{achemso}
\usepackage[utf8]{inputenc}
\usepackage{amssymb,amsmath}
\usepackage{color}
\usepackage{multicol}
\usepackage[version=3]{mhchem}

\usepackage{cancel}
\providecommand{\abs}[1]{\lvert#1\rvert}

\title{Quasi-chemical approximation for polyatomic mixtures}

\author{M. V. D\'avila}
\email{maradav@unsl.edu.ar}
\author{P. M. Pasinetti}
\email{mpasi@unsl.edu.ar}
\affiliation{Departamento de F\'isica, Instituto de F\'isica Aplicada, Universidad Nacional de San Luis-CONICET, Ej\'ercito de los Andes 950, D5700BWS San Luis, Argentina}
\author{D. A. Matoz-Fernandez}
\affiliation{Universit\'e Grenoble Alpes, LIPHY, F-38000 Grenoble, France}
\alsoaffiliation{CNRS, LIPHY, F-38000 Grenoble, France}
\author{A. J. Ramirez-Pastor}
\affiliation{Departamento de F\'isica, Instituto de F\'isica Aplicada, Universidad Nacional de San Luis-CONICET, Ej\'ercito de los Andes 950, D5700BWS San Luis, Argentina}

\begin{document}

\today
\begin{abstract}

The statistical thermodynamics of binary mixtures of polyatomic species was developed on a generalization in the spirit of the lattice-gas model and the quasi-chemical approximation (QCA). The new theoretical framework is obtained by combining: (i) the exact analytical expression for the partition function of non-interacting mixtures of linear $k$-mers and $l$-mers (species occupying $k$ sites and $l$ sites, respectively) adsorbed in one dimension, and its extension to higher dimensions; and (ii) a generalization of the classical QCA for multicomponent adsorbates and multisite-occupancy adsorption. The process is analyzed through the partial adsorption isotherms corresponding to both species of the mixture. Comparisons with analytical data from Bragg–Williams approximation (BWA) and Monte Carlo simulations are performed in order to test the validity of the theoretical model. Even though a good fitting is obtained from BWA, it is found that QCA provides a more accurate description of the phenomenon of adsorption of interacting polyatomic mixtures.
\end{abstract}

\section{Introduction}

The theoretical description of adsorption of polyatomics ranging from small light molecules to hydrocarbons, macromolecules and polymers is a current and exciting topic of research in surface science \cite{Tovbin,Zhdanov,Rudzi,Keller,PRL1}. In practical situations, most adsorbates involved in adsorption experiments are polyatomic in the sense that, when adsorbed, their typical size is larger than the distance
between the nearest-neighbor local minima of the gas–solid potential. This effect, so-called multisite-occupancy adsorption, introduces a high degree of local correlation in the adsorption theories, and it limits the analytical developments in the field of adsorbed polyatomic gases.

The problem becomes even more complicated when the adsorbate is constituted by more than one species. In fact, adsorption of mixtures is a much demanding problem both experimentally and theoretically \cite{Dunne,Li,Smit}. Whereas for pure components the number of adsorbed molecules can be determined accurately by simply measuring the weight increase of the adsorbent sample, for mixtures one has to carry out additional experiments to determine the composition in the sample.

For these reasons, it has been difficult to formulate, in an analytical way, the statistics of occupation for mixtures of correlated particles such as dimers, trimers, etc. In this context, several contributions to the study
of adsorption of polyatomic binary mixtures have been recently introduced by our group \cite{CPL1,CPL4,JCP8,PCCP3}. The first one \cite{CPL1} develops the rigorous statistical thermodynamics of $s$-mer(particle occupying $s$ lattice sites)$-$ $k$-mer(particle occupying $k$ lattice sites) mixtures adsorbed on one-dimensional homogeneous surfaces. The formalism presented in Ref. \cite{CPL1} was the first exact model of adsorption of polyatomic mixtures in zeolites and allows to demonstrate that the adsorption preference reversal (APR) phenomenon \footnote{An unusual feature is observed in the case of methane–ethane mixtures \cite{Abdul,Du}: at low pressure the adsorbed phase is almost entirely ethane, but at higher pressures methane molecules displace ethane molecules. A similar scenario has been observed for different mixtures of linear hydrocarbons in silicalite \cite{Smit,Krishna}, carbon nanotube bundles \cite{Jiang} and metal-organic frameworks \cite{Jiang1}. In all cases, at low (high) pressure the selectivity is toward the larger (smallest) component. This behavior is known as Adsorption Preference Reversal \cite{Ayache}.} is the result of the difference of size (or number of occupied sites) between the adsorbed species. In other words, the results in Ref. \cite{CPL1} revealed that a real description of the phenomenon of APR may be severely misunderstood, if the polyatomic character of the adsorbate is not properly incorporated in the thermodynamic functions from which experiments are interpreted.

In Ref. \cite{CPL4}, the statistical thermodynamics of polyatomic species mixtures adsorbed on two-dimensional substrates was developed on a generalization in the spirit of the lattice-gas model and the classical Guggenheim-DiMarzio approximation \cite{Gugge,Dima}. The theoretical formalism allows to study the problem of adsorption of alkane binary mixtures, leads to the exact solution in one dimension and provides a close approximation for two-dimensional systems accounting multisite occupancy. In Ref. \cite{JCP8}, the multicomponent adsorption of polyatomics was described as a fractional statistics problem, based on Haldane's statistics \cite{Haldane,Wu}. The thermodynamic functions calculated for a monomer–dimer mixture were applied to describe the adsorption of methane–ethane mixtures in zeolites.

Later, a new theoretical approach to treat the statistical thermodynamics of polyatomic species mixtures adsorbed on two-dimensional lattices was presented \cite{PCCP3}. The formalism, based on generalization of the semiempirical approximation for single-component adsorption \cite{LANGMUIR11}, is capable of including the main theories of multisite occupancy adsorption as particular cases. In this framework, a simple adsorption isotherm was obtained by combining exact calculations in one dimension and the Guggenheim-DiMarzio approximation with adequate weights. The reaches and limitations of the theory were analyzed by comparing with Monte Carlo (MC) simulations in square and triangular lattices. The obtained results showed that the semiempirical model represents a significant qualitative advance with respect to existing theoretical models developed to treat the problem of mixture adsorption of long straight rigid rods.

The results in Refs. \cite{CPL1,CPL4,PCCP3} are restricted to the case of non-interacting adsorbates. However, the study of intermolecular forces (adsorbate–adsorbate interactions) is one of the central problems in surface physics and statistical mechanics due, particularly, to the possibility of phase transitions \cite{Stanley,Plischke,Goldenfeld,Yeomans,Patry,Puri} (and the corresponding emerging adsorbate structures that arise on the surface). Among the common types of phase transitions are, condensation of gases, melting of solids, transitions from paramagnet to ferromagnet and order–disorder transitions.

From a theoretical point of view, when nearest-neighbor interactions are present, an extra term in the partition function for interaction energy is required. With this extra term, only partition functions for the whole system can be written. For the one-dimensional lattice, the problem of interacting mixtures with multisite occupancy can be exactly solved and there is no evidence of phase transitions \cite{Manzi}. Close approximate solutions in dimensions higher than one can be obtained, and the two most important of these are the Bragg–Williams approximation (BWA) \cite{Hill} and the quasi-chemical approximation (QCA) \cite{Hill}.

By following this line of reasoning, in Ref. \cite{JCP8} we reported results for interacting mixtures of large molecules adsorbed on two-dimensional lattices. The partition function was written as a product of two contributions, the first being the different ways to array $N_1$ molecules of the species 1, $N_2$ molecules of the species 2, . . . , and $N_m$ molecules of the species $m$ on $M$ homogeneous sites, and the second being a term that takes into account the effect of the adsorbate–adsorbate interactions in the framework of the BWA. The study shows the well-known limitations of a mean-field treatment: prediction of phase transition in a linear lattice, the entropy per site does not depend on the lateral interactions, etc.

The Bragg–Williams approximation is the simplest treatment for interacting adsorbed particles, even in the case of mixtures and multisite occupancy. The present article goes a step further, including the nearest-neighbor interactions by following the configuration-counting procedure of the QCA. For this purpose, a new theoretical formalism is presented based upon (i) the exact analytical expression for the partition function of non-interacting mixtures of polyatomics adsorbed on one dimension and its extension to higher dimensions and (ii) a generalization of the classical QCA in which the adsorbate is a binary mixture of $k$-mers (particle occupying $k$ lattice sites) and $l$-mers (particle occupying $l$ lattice sites). In addition, Monte Carlo (MC) simulations are performed in order to test the validity of the theoretical model. The new theoretical scheme allows us (1) to obtain an approximation that is significantly better than the BWA for mixtures of polyatomics; (2) to reproduce the classical QCA for a binary mixture of monomers \cite{Tovbin} and the exact statistical thermodynamics of interacting mixtures adsorbed in one dimension \cite{Manzi}; (3) to develop an accurate approximation for two-dimensional adlayers accounting for non-ideal gas mixtures and multisite occupancy; and (4) to provide a simple model from which experiments may be interpreted.

The paper is organized as follows: In second section, the model for an interacting polyatomics mixture adsorption is presented. In third section, the exact solution for the one-dimensional problem and the quasi-chemical approximation for polyatomics (QCAPM) adsorbed on a one-dimensional lattice is developed. In addition, the basis of the Monte Carlo simulation scheme in the grand canonical and canonical ensembles is given in the fourth section. The results of the theoretical approach are presented in the fifth section, along with a comparison with Monte Carlo simulation data corresponding to interacting dimers adsorbed on one-dimensional and square lattices. Finally, the conclusions are drawn in last section.

\section{Model}
Let consider a substrate modeled like a regular lattice with connectivity $c$. A binary gas mixture is formed by $k$-mers and $l$-mers which can be adsorbed occupying, respectively, $k$ and $l$ sites arranged linearly on the lattice. Different energies are considered in the adsorption process: (1) $U_k$ ($U_l$), constant interaction energy between a $k$-mer ($l$-mer) unit and an adsorption site, (2) $w_{kl}$ lateral interaction energy between two nearest-neighbor units belonging to a $k$-mer and an $l$-mer (idem for $w_{kk}$ and $w_{ll}$). We denote $N_{kl}$ to the number of $kl$ pairs, in which a $k$-mer's unit is a nearest-neighbor of an $l$-mer's unit (idem for $N_{kk}$ and $N_{ll}$) (see Fig. 1).
\begin{center}
\begin{figure}[hbtp]
\centering
\includegraphics[scale=1.0]{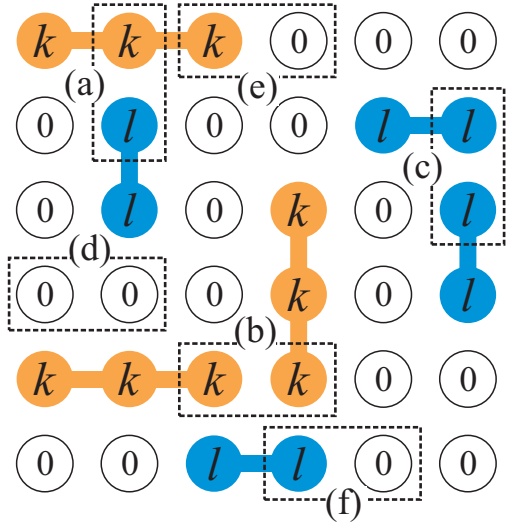}
\caption{Schematic representation of a lattice-gas of dimers ($l=2$, blue circles) and trimers ($k=3$, red circles) adsorbed on a square lattice ($c=4$). The figure shows different types of pairs of sites: a) $kl$, b) $kk$, c) $ll$, d) $00$, e) $k0$ and f) $l0$. }
\end{figure}
\end{center}
The total energy of the system when $N_k$ $k$-mers and $N_l$ $l$-mers are adsorbed keeping a number $N_{kk}$, $N_{ll}$ and $N_{kl}$ of pairs of nearest-neighbors is \begin{equation}
E(N_k,N_l,N_{kk},N_{ll},N_{kl})=N_kkU_k+N_llU_l+N_{kk}w_{kk}+N_{ll}w_{ll}+N_{kl}w_{kl}.
\label{eq1}
\end{equation}

\section{Theory}
\subsection{Exact solution in 1D}

Since the problem can be solved exactly in the special case of monomers mixtures on a one-dimensional lattice and due to two $i$-mers just interact with each other through their ends we can solve the problem of binary mixtures of $i$-mers through an effective lattice.

Let us assume $N_k$ and $N_l$ linear $k-$mers and $l-$mers are adsorbed on a lattice $L$ of M sites, with the lateral interactions explained above. In this lattice, the partial concentrations are given by $\theta_k=kN_k/M$ and $\theta_l=lN_l/M$. We can now mapping $L \rightarrow L'$ from the original lattice $L$ to an effective lattice $L'$, with $M'$ sites, where each empty site of $L$ transforms into an empty one of $L'$, while each set of $x$ sites occupied by an $x$-mer in $L$ is represented by an single $x$-site   in $L'$, like can see in the Fig. 2.
\begin{center}
\begin{figure}[hbtp]
\centering
\includegraphics[scale=0.9]{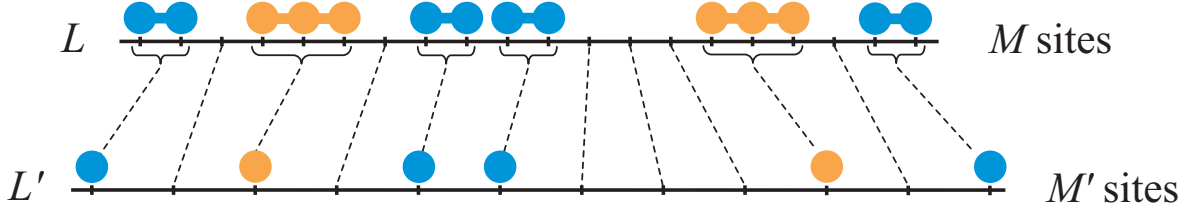}
\caption{Rules for the mapping $L \rightarrow L'$, from the original lattice of polyatomics $L$ to an effective lattice of monomers $L'$.}
\end{figure}
\end{center}
The total number of sites in $L'$ is
\begin{equation}
M'=M-(k-1)N_k-(l-1)N_l
\label{eq0}
\end{equation}
then, the partial concentrations of $L'$ are
\begin{equation}
\theta_x'=\frac{N_x}{M'}=\frac{\theta_x/x}{1-\frac{(k-1)}{k}\theta_k-\frac{(l-1)}{l}\theta_l}
\label{eq01b}
\end{equation}
with $x={k,l}$.
The canonical partition functions $Q(kN_k, lN_l, M, T)$, $Q'(N_k, N_l, M', T)$ in the original and effective lattices must be equal.
Thus
\begin{equation}
\begin{split}
Q(kN_k, lN_l, M, T)=\sum_{\{\Gamma\}}e^{-\beta E(\Gamma)}=Q'(N_k, N_l, M', T)=\sum_{\{\Gamma'\}}e^{-\beta E(\Gamma')}
\label{eq03}
\end{split}
\end{equation}

where $\Gamma$ and $\Gamma'$ refer to a sum over all possible configurations in $L$ and $L'$, respectively, and $\beta=k_B T$, being $k_B$ the Boltzmann constant.

Accordingly, the Helmholtz free energies per site in $L$ and $L'$, $f$ and $f'$, respectively, are related by
\begin{equation}
\begin{split}
\beta f(kN_k, lN_l, M, T)&=-\frac{1}{M}\ln Q(kN_k, lN_l, M, T)\\ &=-\frac{1}{M}\ln Q'(N_k, N_l, M', T)\\ &=\frac{M'}{M}\beta f'(N_k, N_l, M', T)
\label{eq04}
\end{split}
\end{equation}
\begin{equation}
\begin{split}
 \beta f(\theta_k,\theta_l)&=\left[1-\frac{k-1}{k}\theta_k-\frac{l-1}{l}\theta_l\right]\beta f'(\theta_k',\theta_l')
\label{eq06}
\end{split}
\end{equation}
\begin{equation}
\begin{split}
 \beta \mu_x(\theta_k,\theta_l)&=\frac{\partial \beta f}{\partial \theta_x}\\ &=-(x-1)\beta f'(\theta_k',\theta_l')+x\left[1-\frac{k-1}{k}\theta_k-\frac{l-1}{l}\theta_l\right]\frac{\partial\beta f'(\theta_k',\theta_l')}{\partial \theta_x}
\label{eq07}
\end{split}
\end{equation}
where $x=k,l$. Equation \ref{eq07} represents the partial adsorption isotherm expression and it can be calculated using $\beta f'(\theta_k', \theta_l')$ derived from the exact quasi-chemical theory \cite{Hill}.

\subsection{Two-dimensional problem}
The canonical partition function for a two-dimensional system can be written as
\begin{equation}
\begin{split}
Q=&q_k^{N_k}q_l^{N_l}\sum_{N_{kk}}\sum_{N_{ll}}\sum_{N_{kl}}g(N_k,N_l,N_{kk},N_{ll},N_{kl},M)\times \\ &\exp [-\beta(N_kkU_k+N_llU_l+N_{kk}w_{kk}+N_{ll}w_{ll}+N_{kl}w_{kl})],
\label{eq2}
\end{split}
\end{equation}
where $\beta=1/k_BT$, $q_k$ ($q_l$) is the partition function of a single adsorbed particle of the species $k$ ($l$), $N_k$ ($N_l$) is the number of molecules adsorbed on the surface of the species $k$ ($l$), $g(N_k,N_l,M,N_{kk},N_{ll},N_{kl})$ is the number of ways to array $N_k$ $k$-mers and $N_l$ $l$-mers on $M$ sites keeping $N_{kk}$, $N_{ll}$ and $N_{kl}$ pairs of occupied sites.

In a similar way to the QCA for only one species \cite{CQk}, here we calculate the expressions relating $N_{kk}$, $N_{ll}$, $N_{kl}$, $N_{k0}$, $N_{l0}$ and $N_{00}$:
\begin{equation}
ckN_k=2N_{kk}+N_{kl}+N_{0k}+2(k-1)N_k,
\label{eq3}
\end{equation}
\begin{equation}
clN_l=2N_{ll}+N_{kl}+N_{0l}+2(l-1)N_l,
\label{eq4}
\end{equation}
\begin{equation}
c(M-kN_k-lN_l)=2N_{00}+N_{0k}+N_{0l}.
\label{eq5}
\end{equation}

The number of total pairs is
\begin{equation}
\text{Number of total pairs}=\frac{cM}{2}-N_k(k-1)-N_l(l-1),
\label{eq6}
\end{equation}
where ``number of $0k$ pairs'' = ``number of $k0$ pairs'' = $N_{0k}/2$ (the same for number of $l0$ and $kl$ pairs).

The number of ways of assigning a total of $\frac{cM}{2}-N_k(k-1)-N_l(l-1)$ independent pairs to the nine categories $0k, k0, 0l, l0, kl, lk, kk, ll$ and $00$ is
\begin{equation}
\tilde{g}(N_k,N_l,N_{kk},N_{ll},N_{kl},M)=\frac{[\frac{cM}{2}-N_k(k-1)-N_l(l-1)]!}{\left[\left(\frac{N_{ok}}{2}\right)!\right]^2 \left[\left(\frac{N_{ol}}{2}\right)!\right]^2 \left[\left(\frac{N_{kl}}{2}\right)!\right]^2N_{kk}!N_{ll}!N_{00}!}.
\label{eq7}
\end{equation}

This is not equal to $g(N_k,N_l, M, N_{kk},N_{ll},N_{kl})$ in Eq. \ref{eq2}, because treating the pairs as independent entities leads to some non-physical configurations. To take care of this, we assume $g$ proportional to $\tilde{g}$ with a proportionality constant $C(N_k,N_l,M)$
\begin{equation}
g(N_k,N_l,M,N_{kk},N_{ll},N_{kl})=C(N_k,N_l,M)\tilde{g}(N_k,N_l,M,N_{kk},N_{ll},N_{kl})
\label{eq8}
\end{equation}
and
\begin{equation}
\begin{split}
\Omega(N_k,N_l,M)&=\sum_{N_{kk}}\sum_{N_{ll}}\sum_{N_{kl}} g(N_k,N_l,M,N_{kk},N_{ll},N_{kl})\\&= C(N_k, N_l,M) \sum_{N_{kk}}\sum_{N_{ll}}\sum_{N_{kl}}
\tilde{g}(N_k,N_l,M,N_{kk},N_{ll},N_{kl}).
\label{eq9}
\end{split}
\end{equation}
where $\Omega(N_k, N_l, M)$ is the number of ways to arrange $N_k$ $k$-mers and $N_l$ $l$-mers on $M$ sites. There is not an exact expression for $\Omega(N_k, N_l, M)$ in two (or more) dimensions. This is the reason why approximated expressions \cite{PCCP3} have been used to obtain $C(N_k, N_l, M)$. In this work we use an extension of the exact number for one dimension \cite{EA}, because it yields satisfactory results and a simple mathematical equation. $C(N_k, N_l, M)$ is calculated as usual using the maximum-term method \cite{Hill}, replacing the sum in Eq. \ref{eq9} by the maximum term in the sum.

By taking logarithm in Eq. \ref{eq7}, using the Stirling’s approximation and operating, it results
\begin{equation}
\begin{split}
\ln \tilde{g}=&\left[\frac{cM}{2}-N_k(k-1)-N_l(l-1)\right]\ln \left[\frac{cM}{2}-N_k(k-1)-N_l(l-1)\right]-N_{0l}\ln\frac{N_{0l}}{2}\\& -N_{0k}\ln\frac{N_{0k}}{2}-N_{lk}\ln\frac{N_{lk}}{2}-N_{kk}\ln N_{kk}-N_{ll}\ln N_{ll}-N_{00}\ln N_{00}.
\label{eq10}
\end{split}
\end{equation}

From Eqs. \ref{eq3}-\ref{eq5} we find $N_{0k}$, $N_{0l}$ and $N_{00}$, and replace it in Eq. \ref{eq10}, then

\begin{equation}
\begin{split}
\ln \tilde{g}(N_{kk},N_{ll},N_{lk})=&\left[\frac{cM}{2}-N_k(k-1)-N_l(l-1)\right]\ln \left[\frac{cM}{2}-N_k(k-1)-N_l(l-1)\right]\\&-[clN_l-2N_{ll}-N_{lk}-2(l-1)N_l]\ln \left\lbrace \frac{1}{2}\left [clN_l-2N_{ll}-N_{lk}-2(l-1)N_l\right ]\right\rbrace \\&-[ckN_k-2N_{kk}-N_{rk}-2(k-1)N_k]\ln \left\lbrace \frac{1}{2}[ckN_k-2N_{kk}-N_{rk}-2(k-1)N_k]\right\rbrace \\ &-N_{lk}\ln \frac{N_{rk}}{2}-N_{kk}\ln N_{kk}-N_{ll}\ln N_{ll}\\ & -\left [\frac{cM}{2}+N_k(k-ck-1)+N_l(l-cl-1)+N_{ll}+N_{lk}+N_{kk}\right ]\\ & \times \ln \left [\frac{cM}{2}+N_k(k-ck-1)+N_l(l-cl-1)+N_{ll}+N_{lk}+N_{kk}\right ].
\label{eq11}
\end{split}
\end{equation}

In order to maximize $\ln \tilde{g}$ the following procedure is applied

\begin{equation}
\nabla \ln\tilde{g}(N_{kk},N_{ll},N_{lk})=0,
\label{eq13_5}
\end{equation}
\begin{equation}
\begin{split}
\frac{\partial\ln \tilde{g}(N_{kk},N_{ll},N_{lk})}{\partial N_{kk}}=&2\ln\left\lbrace\frac{1}{2} \left[c_k N_k-2N_{kk}-N_{lk}-2(k-1)N_k\right]  \right\rbrace-\ln N_{kk} \\&-\ln \left[\frac{cM}{2}+N_k(k-1-ck)+N_l(l-1-cl)+N_{ll}+N_{kk}+N_{lk}\right]=0,
\label{eq12}
\end{split}
\end{equation}
\begin{equation}
\begin{split}
\frac{\partial\ln \tilde{g}(N_{kk},N_{ll},N_{lk})}{\partial N_{ll}}=&2\ln\left\lbrace\frac{1}{2} \left[c_l N_l-2N_{ll}-N_{lk}-2(l-1)N_l\right]  \right\rbrace-\ln N_{ll} \\&-\ln \left[\frac{cM}{2}+N_l(l-1-cl)+N_l(l-1-cl)+N_{ll}+N_{kk}+N_{lk}\right]=0,
\label{eq13}
\end{split}
\end{equation}
\begin{equation}
\begin{split}
\frac{\partial\ln \tilde{g}(N_{kk},N_{ll},N_{lk})}{\partial N_{lk}}=&\ln\left\lbrace\frac{1}{2} \left[c_l N_l-2N_{ll}-N_{lk}-2(l-1)N_l\right]  \right\rbrace-\ln \frac{N_{lk}}{2} \\&+ \ln\left\lbrace\frac{1}{2} \left[c_k N_k-2N_{kk}-N_{lk}-2(k-1)N_k\right]  \right\rbrace \\&-\ln \left[\frac{cM}{2}+N_k(k-1-ck)+N_l(l-1-cl)+N_{ll}+N_{lk}+N_{kk}\right]=0.
\label{eq14}
\end{split}
\end{equation}

Then
\begin{equation}
N_{kk}N_{ll}=\left (\frac{N_{0k}}{2} \right)^2,
\label{eq19}
\end{equation}
\begin{equation}
N_{ll}N_{00}=\left(\frac{N_{0l}}{2}\right)^2,
\label{eq20}
\end{equation}
\begin{equation}
N_{lk}N_{00}=\frac{N_{0l}N_{0k}}{2}.
\label{eq21}
\end{equation}

We denote the solution of the equations system (Eqs. \ref{eq19}-\ref{eq21}) with the superindex $^*$. From this solution and Eq. \ref{eq9} we can find $C(N_k, N_l, M)$
\begin{equation}
\begin{split}
\Omega(N_k,N_l,M)=&C(N_k, N_l,M) \sum_{N_{kk}}\sum_{N_{ll}}\sum_{N_{kl}}\tilde{g}(N_k,N_l,M,N_{kk},N_{ll},N_{kl})\\& =C(N_k, N_l,M) \tilde{g}^*(N_k,N_l,M,N_{kk}^*,N_{ll}^*,N_{kl}^*).
\label{eq22}
\end{split}
\end{equation}

\begin{equation}
C(N_k, N_l,M)= \frac{\Omega(N_k,N_l,M)}{\tilde{g}^*(N_k,N_l,M,N_{kk}^*,N_{ll}^*,N_{kl}^*)}.
\label{eq23}
\end{equation}
Then
\begin{equation}
g=\frac{\Omega(N_k, N_l, M)\tilde{g}(N_k,N_l,M,N_{kk},N_{ll},N_{kl})}{\tilde{g}^*(N_k,N_l,M,N_{kk}^*,N_{ll}^*,N_{kl}^*)}
\label{eq25}
\end{equation}

or in a simplified form:

\begin{equation}
g=\frac{\Omega\tilde{g}}{\tilde{g}^*}
\label{eq26}
\end{equation}

The partition function can be written as
\begin{equation}
Q=q_k^{N_k}q_l^{N_l}\frac{\Omega}{\tilde{g}^*}\sum_{N_{kk}}\sum_{N_{ll}}\sum_{N_{kl}}\tilde{g}e^{-\beta E},
\label{eq28}
\end{equation}

This sum can be solved applying the maximum-term method again, by means of the following equations
\begin{equation}
N_{kk}N_{00}e^{\beta w_{kk}}=\left (\frac{N_{0k}}{2} \right)^2,
\label{eq29}
\end{equation}
\begin{equation}
N_{ll}N_{00}e^{\beta w_{ll}}=\left(\frac{N_{0l}}{2}\right)^2,
\label{eq30}
\end{equation}
\begin{equation}
N_{lk}N_{00}e^{\beta w_{lk}}=\frac{N_{0l}N_{0k}}{2}.
\label{eq31}
\end{equation}
We denote with superindex $^{**}$ to the variables that solve this equation system, then the Eq. \ref{eq28} is
\begin{equation}
Q=q_k^{N_k}q_l^{N_l}\frac{\Omega\tilde{g}^{**}}{\tilde{g}^*}e^{-\beta E^{**}}.
\label{eq32}
\end{equation}
Taking logarithm
\begin{equation}
\ln Q=N_k\ln q_k+N_l\ln q_l+\ln\Omega+\ln\tilde{g}^{**}-\ln\tilde{g}^*-\beta E^{**}.
\label{eq33}
\end{equation}
The chemical potentials on the adsorbed phase can be calculated from the free energy $F=-\ln Q$,
\begin{equation}
\mu_{k,ads}=\left (\dfrac{\partial F}{\partial N_k}\right )_{N_l},
\label{eq34}
\end{equation}
\begin{equation}
\mu_{l,ads}=\left (\dfrac{\partial F}{\partial N_l}\right )_{N_k},
\label{eq35}
\end{equation}
or
\begin{equation}
\beta\mu_{k,ads}=k\left (\dfrac{\partial \beta f}{\partial \theta_k}\right )_{N_l},
\label{eq36}
\end{equation}
\begin{equation}
\beta\mu_{l,ads}=l\left (\dfrac{\partial \beta f}{\partial \theta_l}\right )_{N_k},
\label{eq37}
\end{equation}
where $f=F/M$ and $\theta_x= xN_x/M$.

In the gas phase the chemical potentials for each species are
\begin{equation}
\beta\mu_{x,gas}=\beta\mu_xº+\ln X_x P,
\label{eq38}
\end{equation}

where $X_x$ is the mole fraction, $\mu_x^0$ is the standard chemical potential
of the $x$-mer:
\begin{equation}
\mu_x^0=-k_BT\ln \left[\left({2\pi m_xk_BT}\over h^2\right)^{3/2}k_BT
\right].
\label{eq40}
\end{equation}

At the equilibrium the chemical potentials are equal in both phases
\begin{equation}
\mu_{x,ads}=\mu_{x,gas},
\label{eq41}
\end{equation}
then
\begin{equation}
\beta\mu_xº+\ln X_x P=x\left (\dfrac{\partial \beta f}{\partial \theta_x}\right )_{N_x},
\label{eq42}
\end{equation}

$N_{kk}^{**}$, $N_{ll}^{**}$, $N_{kl}^{**}$, $\theta_k$ and $\theta_l$ from Eqs. \ref{eq29}-\ref{eq31} and \ref{eq42} ($x=k,l$) were solved using numerical calculations. The data of $\theta_k$ and $\theta_l$ as a function of $\ln P$ are the adsorption isotherms.
\section{Monte Carlo Simulation}
In order to test the applicability of the new theoretical model we perform numerical simulations using hyper-parallel tempering Monte Carlo (HPTMC)\cite{Yan200,Hukushima1996}. The HPTMC method consists in generating a compound system of $R$ non-interacting replicas of the system under study. Each replica is associated with a gas pressure $P_i$, taken from a set of properly selected pressures $\{P_i\}$ \cite{Hartmann2006} \footnote{To determine the number of sampled pressures we used an acceptance ratio of $0.5$ for the swapping move for each pair of replicas.}. Once the values of the gas mixture pressure and molar fractions $X_x$ are set, the chemical potential of each species is obtained an ideal gas mixture, i.e, $\mu_x=\mu^{\rm 0}_x+\ln X_xP$, where $\mu^{\rm 0}_x$  is the standard chemical potential at temperature $T$ and $x=k,l$.

Under these considerations, the simulation process consist in two major subroutines: \textit{replica-update} and \textit{replica-exchange}.

\paragraph{Replica-update.} The adsorption-desorption procedure is as follows: (i) One out of $R$ replicas is randomly selected; (ii) the species $x$ is selected with equal probability from the two species, $k$ and $l$; (iii) a linear $x$-uple of nearest-neighbor sites is selected. Then, if the $x$ sites are empty, an attempt is made to deposit a rod with probability $W_{ads}=\min \lbrace 1,\exp[\beta(\mu_x-\Delta E)]\rbrace$; if the $x$ sites are occupied by units belonging to the same $x$-mer, an attempt is made to desorb this $x$-mer with probability $W_{des}= \min \lbrace1,\exp[\beta(-\mu_x-\Delta E)]\rbrace$ and otherwise, the attempt is rejected. $\Delta E$ is the difference in the configurational energy of the replica between the final and initial states.

\paragraph{Replica-exchange.} Exchange of two configurations $\chi_i$ and $\chi_j$, corresponding to the $i$-th and $j$-th replicas, respectively, is tried and accepted with probability, $W_{accep}(\chi_i\rightarrow \chi_j)=\min \lbrace 1,\exp(\beta\Delta)\rbrace$,  where $\Delta$ in a nonthermal grand canonical ensemble is given by,
\begin{equation}
\Delta=-\left[\left(\mu_k(j)-\mu_k(i)(N_k(j)-N_k(i))\right)+\left(\mu_l(j)-\mu_l(i)(N_l(j)-N_l(i))\right)\right]
\label{eq. Tempering}
\end{equation}

The complete simulation procedure is the following: (1)
replica-update, (2) replica-exchange, and (3) repeat from step (a) $R \times M$ times. This is the elementary step in the simulation process or Monte Carlo step (MCs). Typically, the equilibrium state can be well reproduced after discarding the first $r'=10^6$ MCs. Then, the next $r=10^6$ MCs are used to compute averages.

For each value of pressure $P_i$, the corresponding surface fractions are determinate by simple averages,
\begin{equation}
\theta_x(j)=\frac{1}{r}\sum_{t=1}^{r}\theta_x\left[ \chi_j(t)\right] \qquad\{x = k,l\},
\label{eq. SIN_NUMERO_TODAVIA}
\end{equation}
where $\chi_j(t)$ represents the state of the replica $j$-th at the Monte Carlo time $t$.

\section{Results}

In this section we analyze some examples of QCAPM application. We will consider first the adsorption of two species on a 1D system, and then we will address the adsorption problem in the two-dimensional square lattice. In order to test  the CQAPM clearly, we set $w_{kk}=w_{ll}=w_{kl}=w$. The gas phase is considered as an ideal gas mixture of particles with masses $km_0$ and $lm_0$. The molecule shapes are contemplated only in the adsorbed phase. Then Eq. \ref{eq40} becomes
\begin{equation}
\mu_i^0=-\frac{3}{2}k_BT\ln m_i+C=-\frac{3}{2}k_BT\ln i+C',
\label{eq44}
\end{equation}
where $C$ and $C'$ are constants which can be taken equal to zero without any lost of generality.

Let us consider the adsorption on a 1D system for an equimolar monomer-dimer mixture. Figure 3a) shows the adsorption isotherms for attractive lateral interactions, whereas Fig. 3b) shows the repulsive case. The results have been contrasted with MC data. In both cases, the smallest species fills the monolayer monotonously until completion, while the largest one is present only in a limited range of pressures. In the attractive case as the absolute value of the lateral interaction increases i) the isotherms are shifted towards smaller values of pressure and ii) a greater number of dimers is adsorbed on the surface. The adsorption is more favorable when the interactions are more attractive. The opposite occurs in the repulsive case, as the lateral interaction increases the isotherms are shifted towards larger values of pressure and less dimers are adsorbed. It can be observed in the monomer isotherms that a plateau corresponding to an ordered structure begins to appear, for the strongest interactions cases. It is well known that this is not the case of a phase transition, since it corresponds to a one-dimensional problem. It is worth mention that, as in the 1D case the QCAPM reproduces the exact solution, it would be very useful to apply this approach to alkanes mixtures adsorbed on systems like zeolites (with one-dimensional channels) or nanotubes bundles.

\begin{center}
\begin{figure}[hbtp]
\centering
\includegraphics[scale=0.6, clip=true, trim=0cm 3cm 0cm 1.5cm]{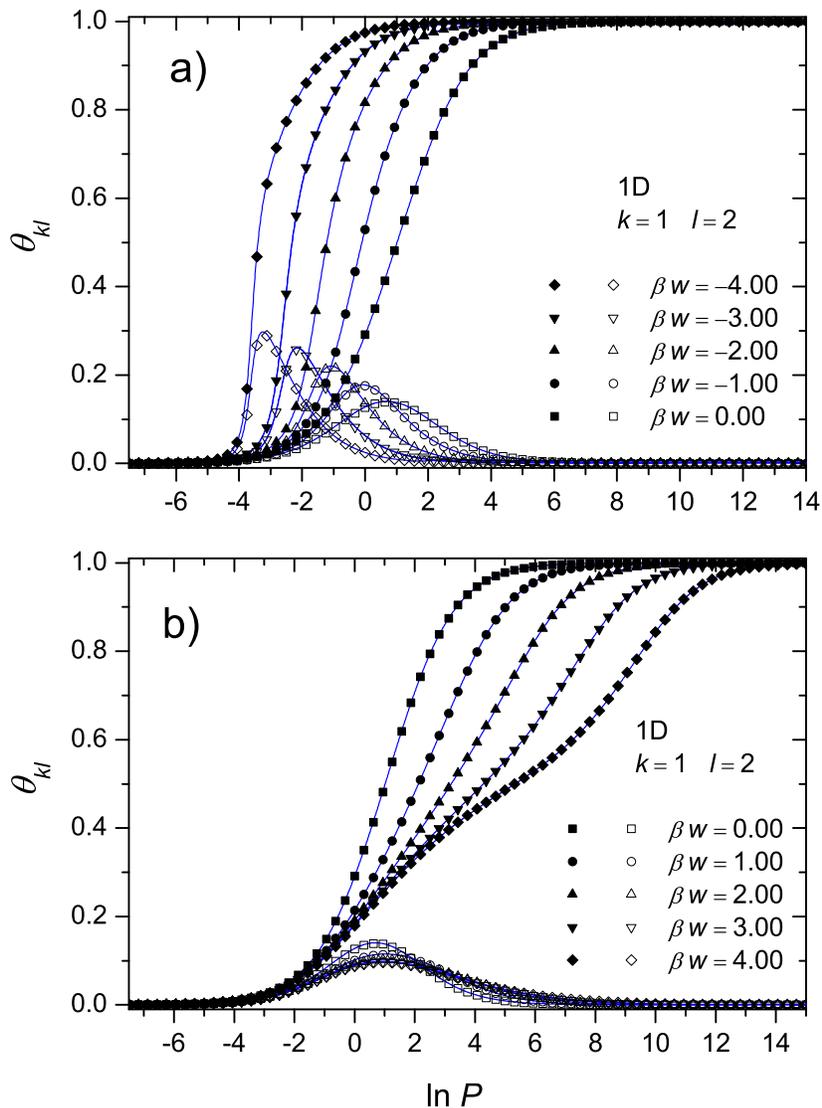}
\caption{Partial adsorption isotherms for the 1D case, $k=1$, $l=2$ and different values of the lateral interaction as indicated. Symbols correspond to MC data whereas lines represent the QCAPM results.}
\end{figure}
\end{center}

We will consider now the square lattice ($c=4$). Figure 4 shows the comparison between the BWA and QCAPM approaches with the MC data, for the case of a monomer-dimer mixture.
For attractive lateral interactions [see Fig. 4a)] both theoretical solutions are, in general, in a good agreement with the simulation data. As in the one-dimensional case, the partial isotherm corresponding to the smallest species increases up to complete the monolayer, whereas the largest one is firstly adsorbed and then is desorbed from the substrate. At high pressures only the monomers are present on the lattice. Figure 4b) shows the case of repulsive lateral interactions where the smallest species presents the characteristic plateau at half coverage as $\ln P$ increases. Regarding the theoretical approaches, in this case the QCAPM shows a much better agreement with the MC data, both qualitative and quantitatively, respect to the BWA. Since the QCAPM is derived from a mean field approximation, it is not capable to reproduces the plateau, despite the higher grade of approximation when compared with the BWA. In order to quantitatively characterize the performance of the theoretical approaches respect to the MC data, we have defined an integral error as:
\begin{equation}
\varepsilon = \int \abs{\theta_{k,\text{theory}}-\theta_{k,\text{MC}}}+\abs{\theta_{l,\text{theory}}-\theta_{l,\text{MC}}} d\ln P
\label{interr}
\end{equation}
Figure 5 shows the curves of this quantity corresponding to the case previously analyzed in Fig. 4. As it can be seen, in all cases the values of QCAPM error are about the half the ones of BWA.

\begin{center}
\begin{figure}[hbtp]
\centering
\includegraphics[scale=0.6, clip=true, trim=0cm 3cm 0cm 1.5cm]{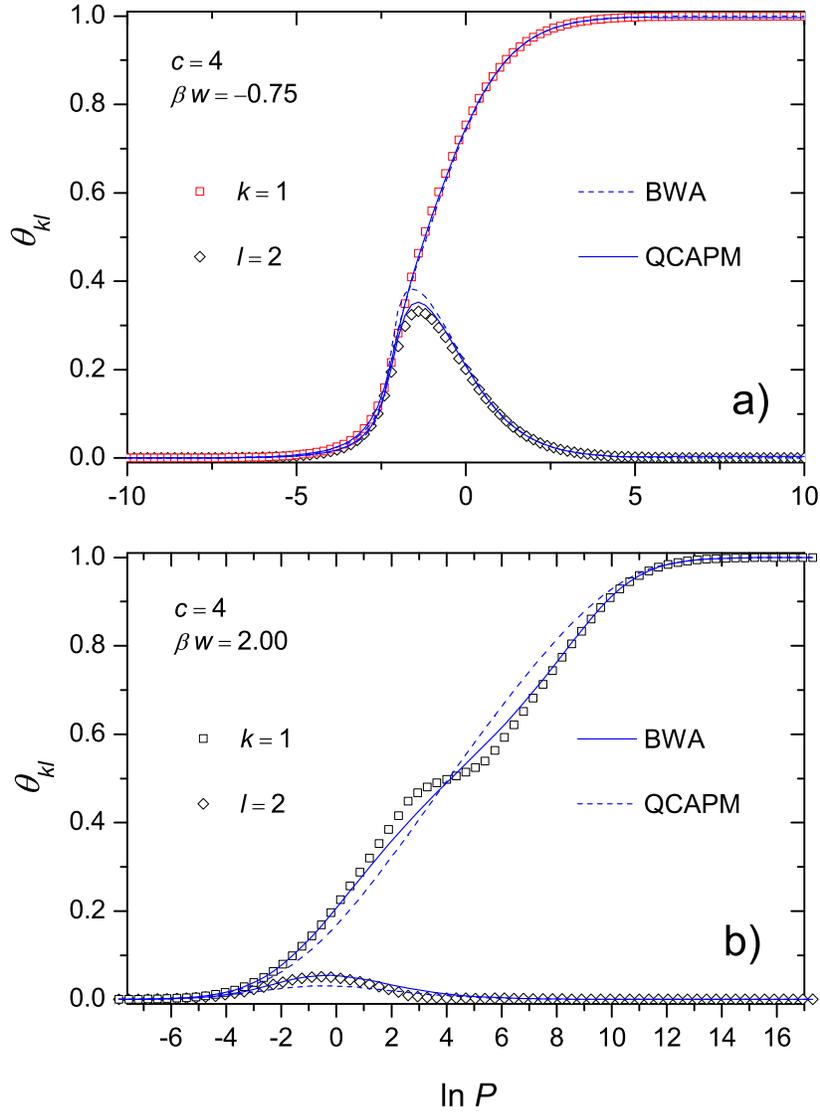}
\caption{Partial adsorption isotherms for the monomer-dimer case on a square lattice with two different values of the lateral interactions: a) $\beta w =-0.75$ (attractive interaction); and b) $\beta w =2.00$ (repulsive interaction). Symbols and lines correspond to MC data and theoretical approaches, respectively.}
\end{figure}
\end{center}

\begin{center}
\begin{figure}[hbtp]
\centering
\includegraphics[scale=0.6, clip=true, trim=0cm 15cm 0cm 1.5cm]{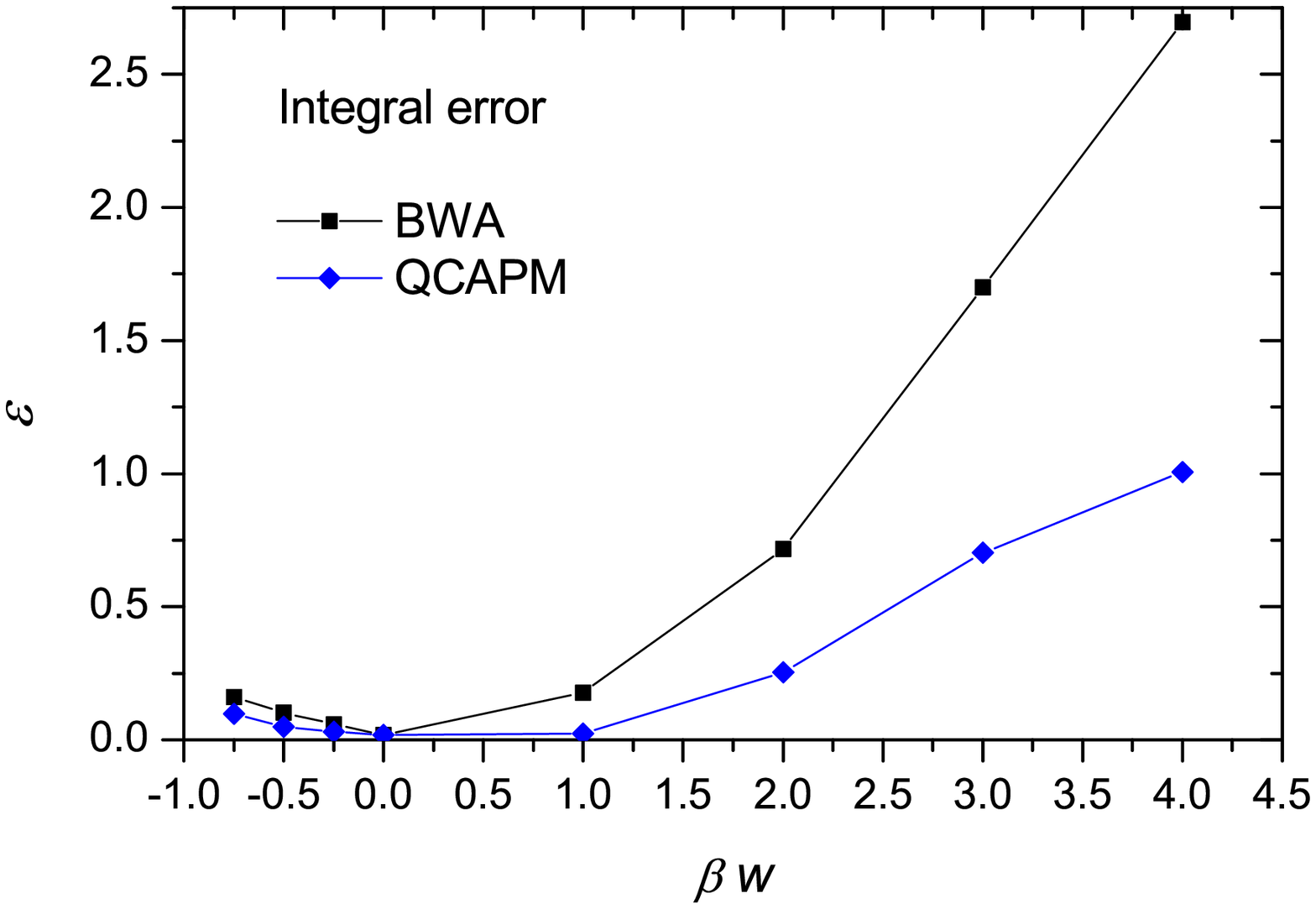}
\caption{Integral error versus lateral interaction energy for the monomer-dimer case shown in Fig. 4.}
\end{figure}
\end{center}

As a way to test further the goodness of the theoretical approaches, Fig. 6 shows the case of a non-equimolar gas mixture, where both attractive and repulsive lateral interactions are considered. The effects of a higher molar fraction on the largest species (Fig. 6 corresponds to $X_l=0.75$) can be seen reflected in the premature growth of the corresponding partial isotherm. However, eventually the largest species is also displaced by the smallest one, as occurs in the equimolar case. Similarly to Fig. 4, the theoretical approaches show a very good agreement in the attractive case. For repulsive interactions the QCAPM clearly shows a better performance over the entire range, except (as expected) in the presence of ordered phases (around $\theta=0.5$ for monomers).

\begin{center}
\begin{figure}[hbtp]
\centering
\includegraphics[scale=0.6, clip=true, trim=0cm 3cm 0cm 1.5cm]{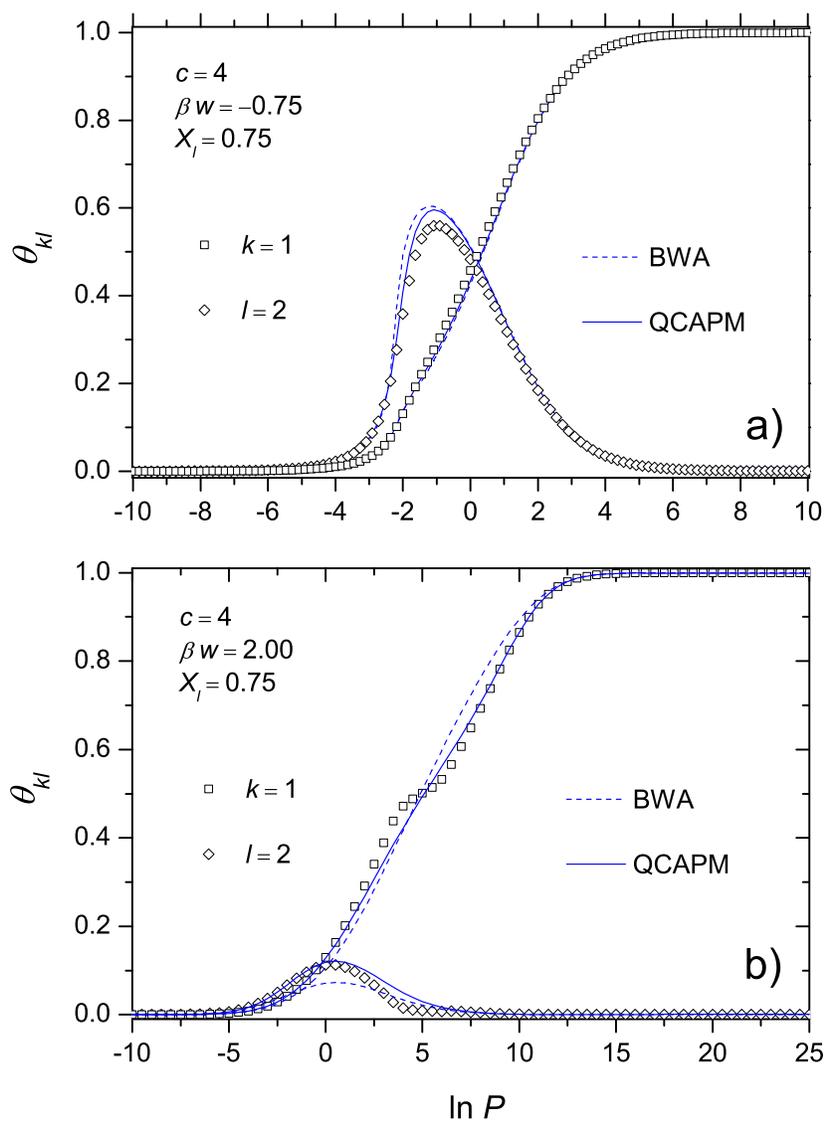}
\caption{Same as Fig. 4 for a non-equimolar mixture. In this case, $X_k=0.25$ and $X_l=0.75$. }
\end{figure}
\end{center}

Finally, Fig. 7 shows the effect of varying the size of the species. Specifically, the case of a mixture of monomer and $l$-mers with repulsive lateral interactions has been considered, for sizes $l$ ranging from 1 to 5. This test is intended in part to stress the goodness of the configurational factor $\Omega(N_k, N_l, M)$, used in the formulation of the QCAPM. In this regard, Fig. 7 shows the integral error, as a function of $l$, both for the MF approximation as for the QCAPM.

\begin{center}
\begin{figure}[hbtp]
\centering
\includegraphics[scale=0.6, clip=true, trim=0cm 15cm 0cm 1.5cm]{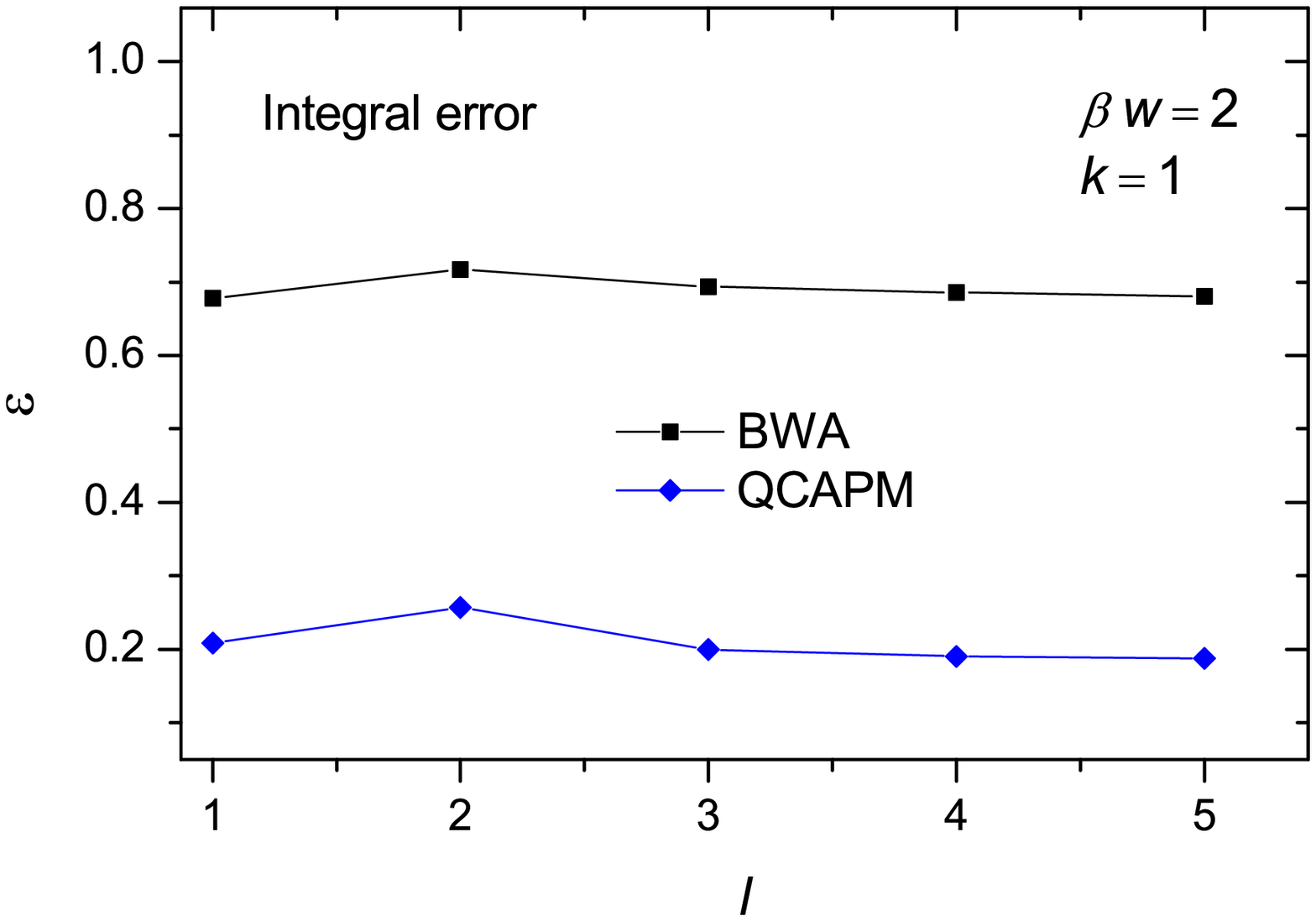}
\caption{Integral error as a function of the species size $l$ for a fixed value of $k$. In the case of the figure, $k=1$, $X_k=X_l=0.5$, $\beta w =2.00$ and $c=4$.}
\end{figure}
\end{center}

\section{Conclusions}

In the present work we have addressed the adsorption problem of a interacting polyatomics mixture by following (i) a generalization of the classical QCA configuration-counting procedure, (ii) the exact analytical expression for the configurational factor of polyatomics mixtures adsorbed in one dimension and its extension to higher dimensions.

Firstly the one-dimensional case was tested with the exact solution for a polyatomics binary mixture obtained from an effective lattice and the exact solution of monomeric mixture adsorption.

Two-dimensional case was also analyzed for adsorption on a square lattice considering both  attractive and repulsive lateral interactions. Additionally, different molar fractions were explored observing the good performance of the theory. On the other hand, Monte Carlo simulations were performed in order to test the validity of the theoretical model. QCAPM and BWA were contrasted with the simulation data qualitative and quantitative way by mean an error measure. In all the cases QCAPM was the best approximation, as expected.

In order to improve the theory developed here, it would be interesting to test another different configurational factors obtained from the main theoretical models developed to treat the polymer adsorption problem\cite{IJMPB1}.

\section{Acknowledgement} This work was supported in part by CONICET (Argentina) under project number PIP 112-201101-00615; Universidad Nacional de San Luis (Argentina) under project 322000 and the National Agency of Scientific and Technological Promotion (Argentina) under project PICT-2013-1678.

\renewcommand{\refname}{\section{References}}

\bibliographystyle{plain}
\bibliography{biblio}
\newpage
\section{Figure captions}

Fig. 1: Schematic representation of a lattice-gas of dimers ($l=2$, blue circles) and trimers ($k=3$, red circles) adsorbed on a square lattice ($c=4$). The figure shows different types of pairs of sites: a) $kl$, b) $kk$, c) $ll$, d) $00$, e) $k0$ and f) $l0$.

Fig. 2: Rules for the mapping $L \rightarrow L'$, from the original lattice of polyatomics $L$ to an effective lattice of monomers $L'$.

Fig. 3: Partial adsorption isotherms for the 1D case, $k=1$, $l=2$ and different values of the lateral interaction as indicated. Symbols correspond to MC data whereas lines represent the QCAPM results.

Fig. 4: Partial adsorption isotherms for the monomer-dimer case on a square lattice with two different values of the lateral interactions: a) $\beta w =-0.75$ (attractive interaction); and b) $\beta w =2.00$ (repulsive interaction). Symbols and lines correspond to MC data and theoretical approaches, respectively.

Fig. 5: Integral error versus lateral interaction energy for the monomer-dimer case shown in Fig. 4.

Fig. 6: Same as Fig. 4 for a non-equimolar mixture. In this case, $X_k=0.25$ and $X_l=0.75$.

Fig. 7: Integral error as a function of the species size $l$ for a fixed value of $k$. In the case of the figure, $k=1$, $X_k=X_l=0.5$, $\beta w =2.00$ and $c=4$.
\newpage
\section{Graphical TOC Entry}

\begin{center}
\begin{figure}[hbtp]
\centering
\includegraphics[scale=0.8]{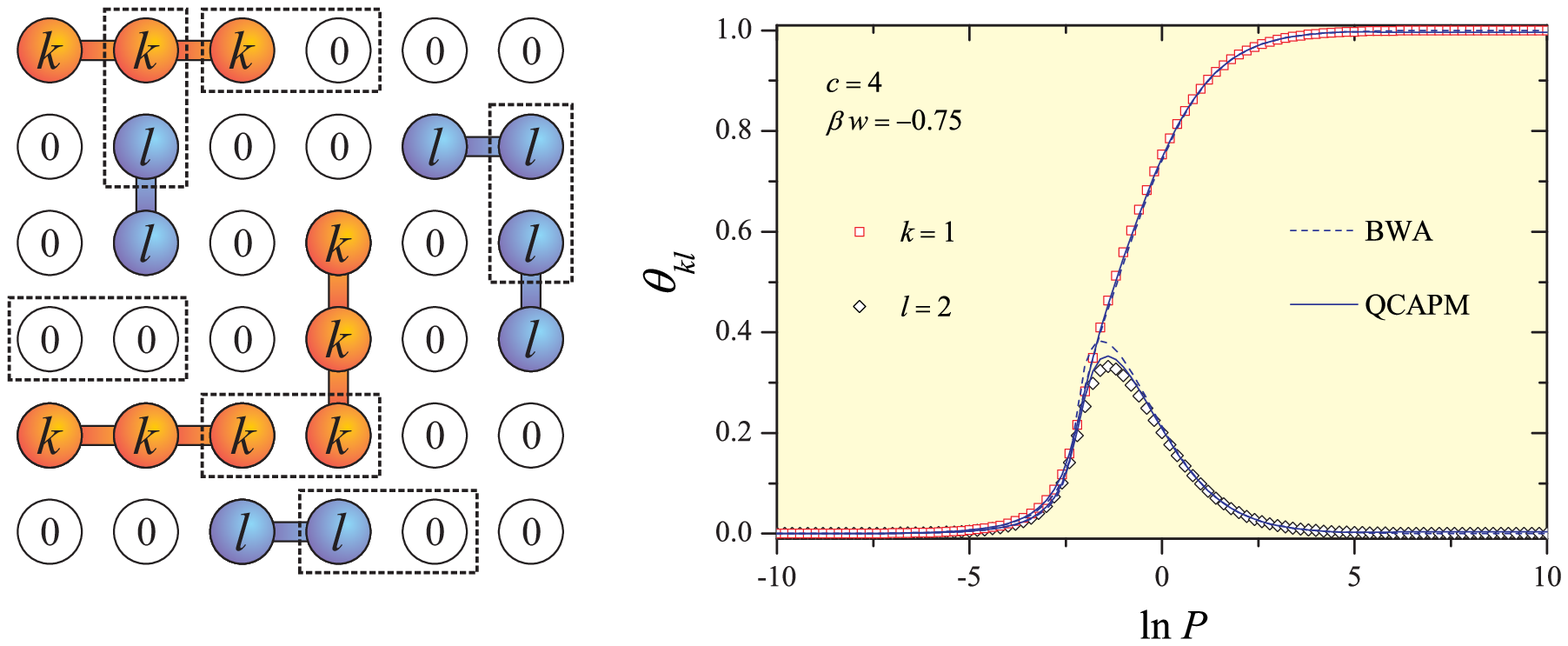}
\end{figure}
\end{center}

\end{document}